\documentclass[a4paper,fleqn]{cas-dc}

\usepackage[authoryear]{natbib}

\usepackage{xcolor}

\def\tsc#1{\csdef{#1}{\textsc{\lowercase{#1}}\xspace}}
\tsc{WGM}
\tsc{QE}
\tsc{EP}
\tsc{PMS}
\tsc{BEC}
\tsc{DE}

\begin{document}
\let\WriteBookmarks\relax
\def\floatpagepagefraction{1}
\def\textpagefraction{.001}
\shorttitle{IMFIT-MW: Multiwavelength photometric decomposition with IMFIT}
\shortauthors{I.V. Chugunov}

\title [mode = title]{IMFIT-MW: Multiwavelength photometric decomposition with IMFIT}                    

\author[1,2]{Ilia V. Chugunov}[orcid=0009-0005-4243-6368]
\ead{21ilya07@gmail.com}

\affiliation[1]{organization={Pulkovo Astronomical Observatory, Russian Academy of Sciences},
                addressline={Pulkovskoye Chaussee 65/1}, 
                city={St. Petersburg},
                postcode={196140},  
                country={Russia}}
                
\affiliation[2]{organization={Sternberg Astronomical Institute, Lomonosov Moscow State University},
	addressline={Universitetsky Pr. 13}, 
	city={Moscow},
	postcode={119234},
	country={Russia}}

\credit{The work was done by the single author}

\begin{abstract}[fragile]
	
Photometric decomposition is often used to study galaxy structure by modelling surface brightness distribution in its components. However, most of the relevant codes are designed for single-band analysis.

\noindent I present \texttt{IMFIT-MW}: a lightweight, open source Python tool that extends \texttt{IMFIT} capabilities to simultaneously model multiwavelength images. \texttt{IMFIT-MW} is able to set shared parameters between filters and to impose smooth variation of them with wavelength. \texttt{IMFIT-MW} achieves this by using \texttt{IMFIT} to produce model brightness distributions for given parameters, whereas hyperparameters defining their wavelength dependence are fitted by an external optimizer. The code is available on GitHub (\url{https://github.com/IVChugunov/IMFIT-MW}).

\noindent I provide a few examples of use, showing that \texttt{IMFIT-MW} matches \texttt{GALFITM} in single-thread performance and can have advantage of multithreading. In a broader sense, I demonstrate the possibility to extend \texttt{IMFIT} features even without modifying its source code.
\end{abstract}



\begin{keywords}
galaxies: photometry\\
galaxies: structure\\
methods: data analysis
\end{keywords}

\maketitle

\section{Introduction}

Photometric decomposition is a powerful tool to retrieve structural parameters of galaxies and their components. This method implies that the surface brightness of each component of a galaxy can be represented by some analytical function, and galaxy itself can be treated as a sum of its components. The parameters of model components are considered to be representative of their real counterparts, and thus decomposition is focused on finding best-fitting model parameters, being essentially an optimization problem.

In a photometric decomposition framework, there are multiple analytical functions commonly used to describe surface brightness of galaxy components. Functional form may be common to different examples of galaxies (for example, being used to describe bars in galaxies), but exact parameters vary from one to another. For example, one can consider the S{\'e}rsic function~\citep{Sersic1963}. It may be used to describe either the entire galaxy, or only its bulge or other component and to be accompanied by other functions, but it does not matter now:

\begin{equation}
	\label{eq:Sersic}
	I(r) = I_e \exp \left\{-b_n \left[\left(\frac{r}{r_e}\right)^{1/n} - 1\right]\right\}
\end{equation}

The radial surface brightness distribution $I(r)$ here is governed by three parameters: $r_e$ is a half-light radius containing half of luminosity, $I_e$ is effective brightness and $n$ is S{\'e}rsic index defining the overall shape of the profile. $b_n$ is defined solely by $n$, therefore it is not an independent parameter. this coefficient exists to ensure $I(r_e) = I_e$ by definition of $I_e$. To obtain the actual surface brightness distribution on 2D image, one also needs the coordinates of its center, and to account for the possible isophote flattening, their positional angle and ellipticity are used. These values are also counted as four more parameters, and together they are optimized during the fitting process. Multi-component models (bulge + disc is one of the simplest examples) usually feature other functions which are commonly associated with specific structural components of galaxies. In particular, discs can be modelled with exponential radial distribution~\citep{Freeman1970}, which may feature a break, i.e. an abrupt change of exponential scale at some radius~\citep{Pohlen2008}. For bars, Ferrers function~\citep{Ferrers1877} is often used~\citep{Williams2017}, but S{\'e}rsic profile may be applied as well~\citep{Erwin2023}. In any case, each component introduces some photometric parameters associated with it to model.

One may be interested in multiwavelength structure and properties of a galaxy. It seems straightforward to do decomposition separately for images in each filter, but this approach essentially misses the information that all images belong to the same galaxy. For example, some parameters are expected to be constant (like the position of galactic centre or disc orientation) or vary only a little and smoothly from band to band. It may be often reasonable to assume their linear or low-degree polynomial variation with filter wavelength or its logarithm. For instance, one can compare the values of different photometric parameters in multiwavelength studies such as \citet{Casasola2017, Gong2023, Marchuk2024}; also see \citet{Vika2013, Casura2022, Haussler2022}.

If images are fitted one by one, there is no way to establish such a relation --- at most, it is possible to fix all parameters to some pre-defined value. The only way to reliably and consistently impose such constraints requires fitting all images simultaneously using a different parametrization, having fitted parameters define per-band photometric quantities.

Among the most popular programs for decomposition, there are \texttt{GALFIT}~\citep{Peng2002, Peng2010}, \texttt{BUDDA}~\citep{deSouza2004}, \texttt{GIM2D}~\citep{Simard2002}, \texttt{IMFIT}~\citep{Erwin2015} and others, each one having its own advantages. However, decomposition routines are predominantely designed for fitting a single image. Among well-known codes, the support of multiwavelength fitting is much less common. For example, \texttt{GALFITM}~\citep{Vika2013, Haussler2022} essentially optimizes not single-band parameters, but a set of hyperparameters which define polynomial dependence of each parameter on wavelength. This program is based on \texttt{GALFIT} and retains its functions and features. Another example is \texttt{ProFit}~\citep{Robotham2017}: a Bayesian fitting tool, newer versions of which also support multiwavelength fitting~\citep{Casura2022}. Both \texttt{GALFITM} and \texttt{ProFit} are also incorporated into automated pipelines \texttt{MegaMorph}~\citep{Haussler2013} and \texttt{ProFuse}~\citep{Robotham2022}, respectively.

However, for some problems, features present in specific programs are important, and thus independent, problem-specific implementations of multiwavelength fitting exist, such as in~\citet{Tsukui2026}. For example, \texttt{IMFIT} is open-source and allows one to easily add custom models to use them in decomposition. Consequently, \texttt{IMFIT} was selected to conduct studies where X-structures in edge-on galaxies~\citep{Smirnov2020, Smirnov2026} and spiral arms~\citep{Chugunov2024, Chugunov2025a} were explicitly modelled. Another advantage of \texttt{IMFIT} is the ability to perform multiprocessing. Therefore, the extension of other packages for decomposition to handle multiwavelength models is useful, and in this work I aim to create the extension of \texttt{IMFIT} capable of solving this problem, which I called \texttt{IMFIT-MW}.

\section{Method and implementation}

\subsection{Multiwavelength parametrization}

The general idea underlying the method is described in~\citet{Vika2013}, relevant to \texttt{GALFITM} implementation. I will describe the method using the following example. Suppose one aims to do S{\'e}rsic fitting of five images in $ugriz$ filters and wants to establish the linear dependence of S{\'e}rsic index $n$ on wavelength $\lambda$. If single-band fitting is done for each individual image, one obtains five independent values of $n_{ugriz}$ which are not guaranteed to follow the linear dependence on $\lambda$ and may have outliers. Instead, one should set up a S{\'e}rsic function which parameters depend on wavelength and fit wavelength-dependent data, i.e. essentially images in all filters simultaneously. One may define $n(\lambda) = n_0 + n_1 \lambda$, and therefore wavelength-dependent S{\'e}rsic function will be parametrized by two parameters $n_0$ and $n_1$ instead of five $n_{ugriz}$.

Speaking more generally, this approach requires to introduce wavelength dependence into a function that defines brightness distribution. This is done by adding said dependence to its parameters $p(\lambda)$ which, in turn, is defined by another set of quantities. Hereafter, I will call these quantities as hyperparameters: they define the values of ordinary photometric parameters (hereafter simply ``parameters'') at all wavelengths under discussion. Therefore, wavelength-dependent brightness distribution can be redefined to ultimately depend only on hyperparameters. To distinguish between parameter value in some filter and the set of values at at all filters, I refer to them as to single-band parameters and wavelength-dependent parameters respectively, where this is relevant.

In the example above, $n$ is wavelength-dependent parameter, with five values $n_{ugriz}$ are single-band parameters. Two linear coefficients $n_0$ and $n_1$ are hyperparameters. Note that, even for a given form of $p(\lambda)$, like a linear function, one can select different set of hyperparameters. For example, wavelength dependence in \texttt{GALFITM} is defined as a polynomial of required degree, and hyperparameters are the coefficients of Chebyshev polynomial expansion.

\texttt{IMFIT-MW} similarly adopts polynomials to characterize wavelength dependence. If one deals with $N$ wavelengths, the polynomial may have any degree $0 \leq M \leq N - 1$ for any wavelength-dependent parameters. For each of them, $M = 0$ makes it constant, $M = N - 1$ means that its corresponding single-band parameters are essentially independent from each other, and any other $M$ allows them to be different, but not independent from each other. The number of hyperparameters is therefore $M + 1$. This part is similar to \texttt{GALFITM}; however, the selection of hyperparameter differs. \texttt{IMFIT-MW} uses single-band parameters at $M + 1$ filters (which are called control filters); they are sufficient to define a polynomial unambiguously. The example of S{\'e}rsic fitting discussed above corresponds to $M = 1$ and $n$ must have 2 control filters out of 5; user can select any possible pair. For example, if $u$ and $z$ are control filters, then $n_u$ and $n_z$ treated as hyperparameters and fitted; they define the remaining $n_g, n_r, n_i$. Such a parametrization was selected because all hyperparameters in this case correspond to physically meaningful quantities, which means it becomes easier to control fitting process and becomes possible to impose band-specific constraints straightforwardly.

\subsection{Additional features}

The polynomial representation of wavelength dependence is not hard-coded. User may add their own function to configuration file, if needed. This may be justified if low-degree polynomial cannot adequately describe the wavelength dependence: for example, if some images belong to other part of the spectrum, like MIR or UV compared to optical, or one also deals with images in emission lines such as H$\alpha$.

Additionally, the user can set some photometric parameters to share their value with another. For example, one can model edge-on galaxies with bulge + disc model and guarantee that position angles of both structures are equal while being fitted, effectively ensuring that bulge oblateness is aligned with the axis of a galaxy. This feature may be useful even regardless of multiwavelength fitting.

\subsection{Implementation architecture}

In any model optimization, with photometric decomposition is no exception, model parameters are improved iteratively. They used to obtain weighted residuals from data, and model parameters are updated accordingly to get smaller residuals at the next iteration. In common decomposition codes, data is a single 2D image, and model parameters are ordinary photometric parameters. Therefore, the primary technical change which had to be done in \texttt{IMFIT-MW} compared to basic \texttt{IMFIT} is to make it handle multiple 2D images at once and use hyperparameters as model parameters.

I decided not to modify \texttt{IMFIT} source code, but to create \texttt{Python} code to selectively access existing features of \texttt{IMFIT} and supplement them with some external tools instead. This approach demonstrates the advantages of \texttt{IMFIT}: its modularity and flexibility allow one to implement simple solutions for the problems which initially were out of the scope of original program.

\begin{figure}
	\centering
	\includegraphics[width=.9\columnwidth]{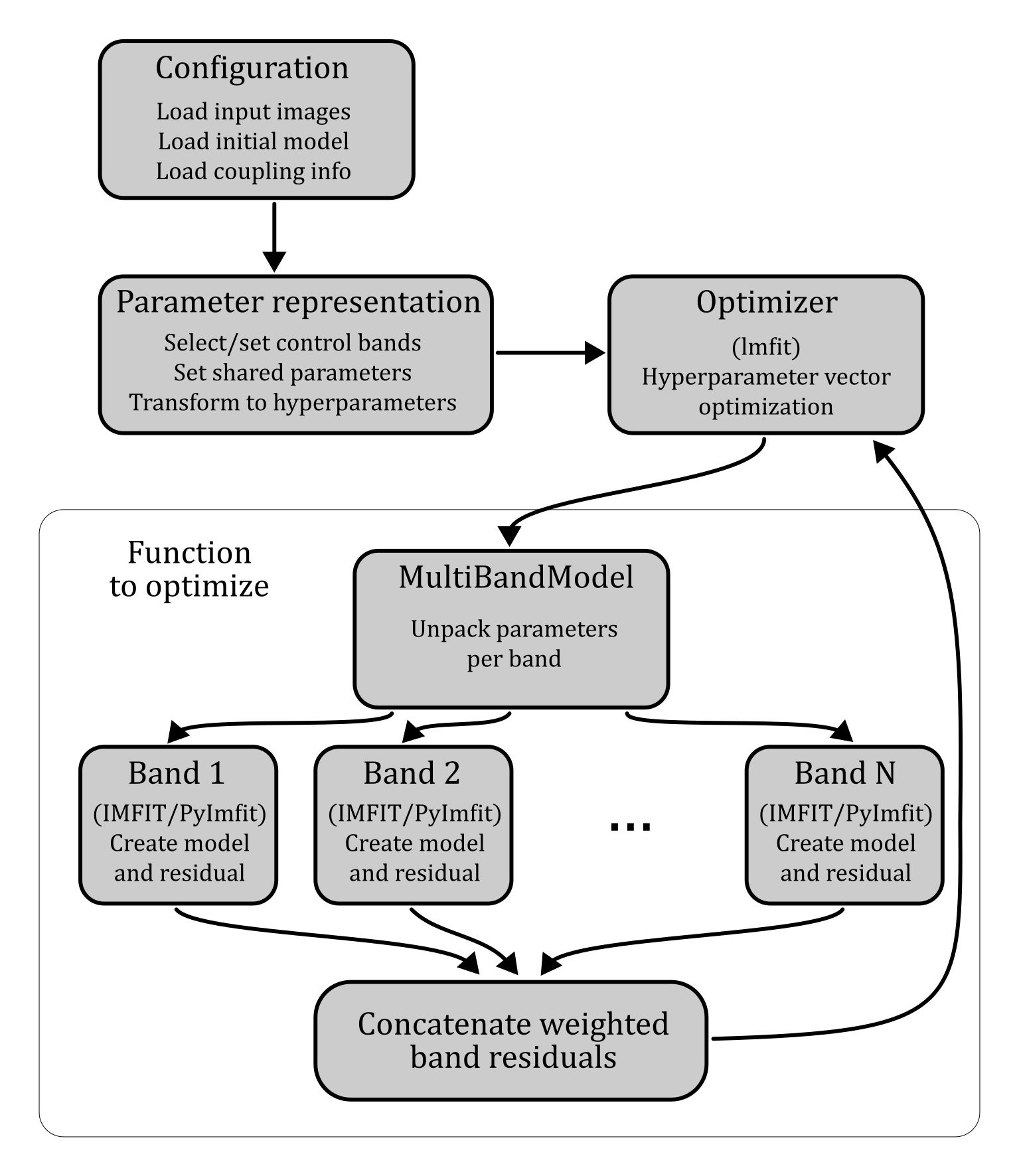}
	\caption{A schematic workflow of \texttt{IMFIT-MW}.}
	\label{fig:workflow}
\end{figure}

Specifically, \texttt{IMFIT-MW} uses \texttt{IMFIT} (via \texttt{PyImfit} interface) for producing model brightness distributions from given parameters in each band, including convolution with PSF. The process of hyperparameter optimization is handled by \texttt{lmfit} library: different optimizers are available, and may be selected by user. The remaining part of \texttt{IMFIT-MW} mostly handles the communication between \texttt{IMFIT} and \texttt{lmfit}: it transforms hyperparameters into single-band values, calculates residuals and transfers them back into optimizer at each iteration. Note that optimizer deals with the unified weighted residual vector, which is created by concatenating per-band weighted residuals. No additional weighting between different bands is needed: individual pixels are weighted according to relevant noise maps. Also, it processes inputs and final outputs as well. In Fig.~\ref{fig:workflow}, I present a schematic workflow of \texttt{IMFIT-MW}. The program is open-source: the code with example of use is openly available on GitHub (\url{https://github.com/IVChugunov/IMFIT-MW}).

\subsection{\texttt{PyImfit} modifications}
\label{sec:PyImfit}
As it was mentioned, \texttt{IMFIT-MW} requires \texttt{PyImfit} library; although the standard version works, it is preferable to install \texttt{PyImfit} with minor modifications available on GitHub (\url{https://github.com/IVChugunov/pyimfit}). The issue with standard \texttt{PyImfit} is that it goes with bundled old version of \texttt{FFTW} and during installation it is linked by default. As \texttt{IMFIT} itself requires \texttt{FFTW}, this seems redundant and, more importantly, may lead to slower convolution compared to \texttt{IMFIT} if it was compiled with a newer version. Second, standard \texttt{PyImfit} relies on \texttt{Cython}, and by default GIL prevents per-band multithreading. With these issues addressed in modified \texttt{PyIMFIT}, performance is increased and multithreading is available.

Note that \texttt{PyImfit} installation from source rather than from \texttt{PyPI} is the only way to use custom \texttt{IMFIT} functions. For this, one should first add relevant functions into \texttt{IMFIT} embedded into source code of \texttt{PyImfit} in an ordinary way and install \texttt{PyImfit} from source (see GitHub repository for instructions).

\section{Demonstration}

\subsection{Performance}

Photometric decomposition is time-consuming, and it is important to assess code performance, so I made two benchmarks of different workload to fit multiwavelength images of a galaxy. In particular, I compare \texttt{IMFIT-MW} with its natural competitor \texttt{GALFITM}. Also, I run \texttt{IMFIT-MW} applied to single-band image: this was made to compare it with \texttt{IMFIT} and \texttt{GALFIT} to check how significant the overhead that \texttt{IMFIT-MW} adds, and also check the consistency between parameters obtained with multiwavelength and single-band fitting (see Section~\ref{sec:validation}).

\begin{figure}
	\centering
	\includegraphics[width=.9\columnwidth]{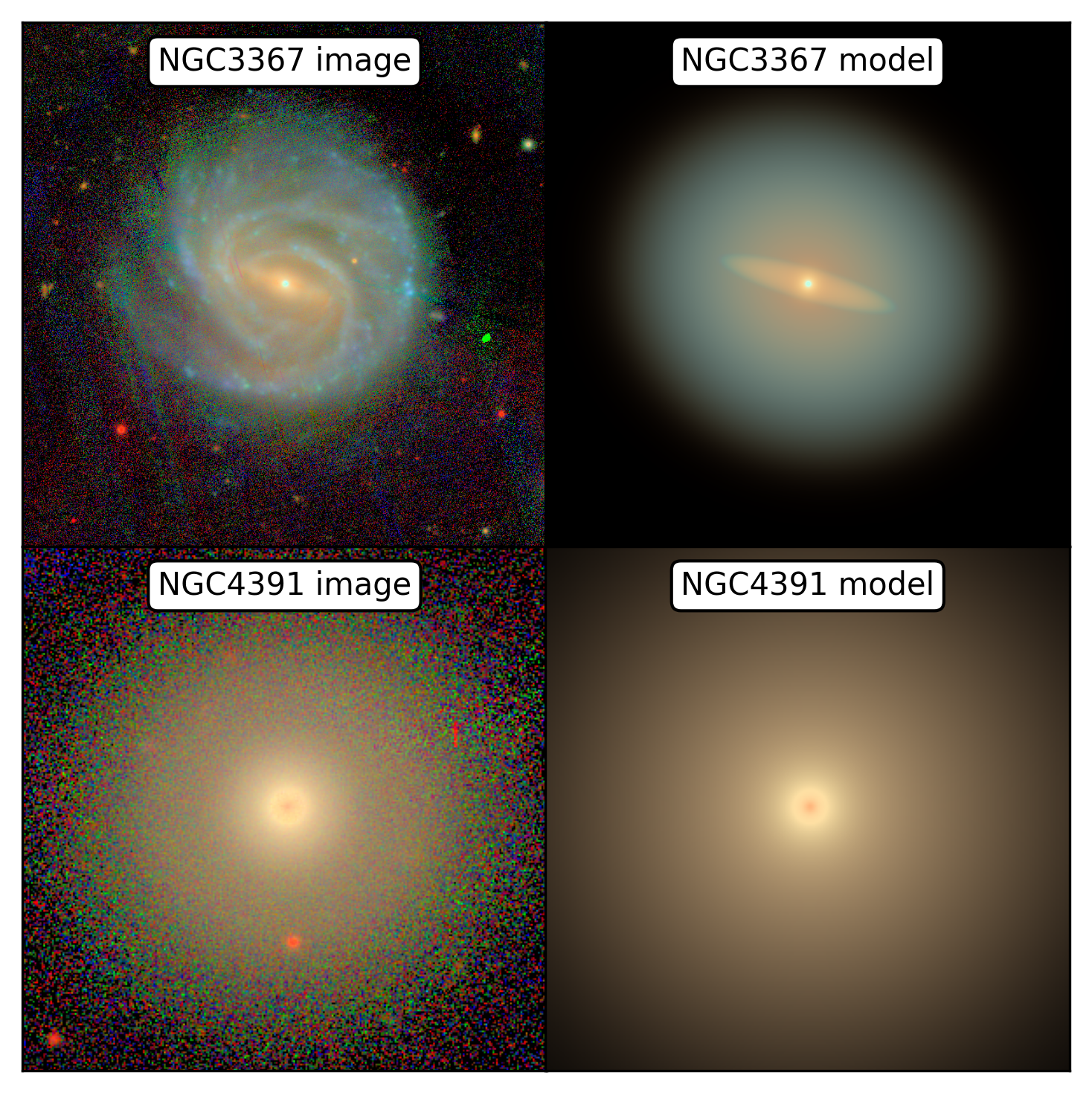}
	\caption{Images of NGC~3367 and NGC~4391 and their multiwavelength models obtained with \texttt{IMFIT-MW}. RGB pictures are produced with $z$, $r$ and $g$ images using~\citet{Lupton2004} scaling.}
	\label{fig:generic_models}
\end{figure}

Benchmarks include imaging data, including noise image and PSF, as well as initial guess created manually. The properties of both galaxies relevant to benchmarking are listed in Table~\ref{tab:benchmark}. I used imaging data from Pan-STARRS; Two galaxies used for benchmarking are NGC~4391 and NGC~3367. The first galaxy is S0 of roughly 1.2' in diameter, so its cutout is relatively small and I adopted a simple model consisting of S{\'e}rsic bulge and exponential disc. Other galaxy is larger (2' in diameter) and has more complex morphology (it is SBc), so use more complex model to fit it, consisting of S{\'e}rsic bulge, Ferrers bar, S{\'e}rsic disc and unresolved point source in the centre. Although there would be more physical sense in modelling disc of NGC~3367 with a broken exponential function (and profile breaks are common features in disc galaxies, see~\citealt{Pohlen2008}), it does not exist as a stand-alone function in \texttt{GALFIT}, so I used S{\'e}rsic instead to reproduce brightness plateau at the centre. To further increase the complexity difference between benchmarks, I use 5 filters for NGC~3367 and 4 for NGC~4391. I set geometric parameters (location, position angle and ellipticity) to be constant across bands, whereas brightness-related parameters were independent, and parameters related to brightness distribution shape (such as radii and S{\'e}rsic $n$) are allowed to vary linearly. In Fig.~\ref{fig:generic_models}, I show these galaxies along with their multiwavelength models obtained with \texttt{IMFIT-MW}.

The selection of models in \texttt{GALFIT} and \texttt{IMFIT} differs as well as the definition of photometric parameters. Most importantly, \texttt{GALFIT} functions are parametrized by either surface brightness expressed in mag/arcsec$^2$ or by integrated magnitude, whereas \texttt{IMFIT} deals with image units directly. Therefore, I selected function sets that can be produced in both programs. While making initial configuration files for \texttt{GALFITM} and \texttt{IMFIT-MW}, I ensured they both produce similar model images, so neither program had an advantage of its initial guess being better.

Benchmarking was done at my personal computer with Intel Core i7-12650H CPU (2022). \texttt{IMFIT-MW} includes \texttt{IMFIT} of version 1.9.1 and \texttt{lmfit} of version 1.3.4. \texttt{GALFITM} is 1.4.4 and based on \texttt{GALFIT} 3.0.5. The resulting time measurements are presented in Table~\ref{tab:benchmark}. All IMFIT-MW benchmarks reported here use the modified \texttt{PyImfit} version described in Section~\ref{sec:PyImfit}. The modifications are not required for correctness, but improve performance. Note that the configuration files required to run these benchmarks are available on GitHub (\url{https://github.com/IVChugunov/IMFIT-MW}).

\begin{table}[width=.9\linewidth,cols=3,pos=h]
\caption{Benchmark characteristics and performance comparison. Single-band benchmarks stand for fitting a single image using the model with the same set of components. Naturally, fitting multiple images, even independently, requires proportionally more time.}\label{tab:benchmark}
\begin{tabular*}{\tblwidth}{@{} l|rr@{} }
\toprule
& \multicolumn{2}{c}{Benchmark} \\
& NGC 4391 & NGC~3367 \\
\midrule
Filters & 4 ($griz$) & 5 ($grizy$)\\
Image size (px) & 300 $\times$ 300 & 800 $\times$ 800 \\
PSF size (px) & 51 $\times$ 51 & 51 $\times$ 51 \\
Model & S+E & S+F+S+P \\
N wav.-dep. & \multirow{2}{*}{9} & \multirow{2}{*}{15} \\
parameters & & \\
N fitted & \multirow{2}{*}{18} & \multirow{2}{*}{36} \\
parameters & & \\
\midrule
Code & \multicolumn{2}{c}{Time, s} \\
\midrule
\texttt{IMFIT-MW} & Overall: 25.5 & 536 \\
(multithreading) & CPU: 113 & 1708 \\
\texttt{IMFIT-MW} & \multirow{2}{*}{85.0} & \multirow{2}{*}{1507} \\
(single-thread) & & \\
\texttt{GALFITM} & 210 & 1477 \\
\midrule
& \multicolumn{2}{c}{Single-band benchmarks} \\
\midrule
\texttt{IMFIT-MW} & 6.8 & 76.5 \\
\texttt{IMFIT} & Overall: 1.7 & 34.9 \\
(multithreading) & CPU: 16.7 & 262 \\
\texttt{IMFIT} & \multirow{2}{*}{5.2} & \multirow{2}{*}{119} \\
(single-thread) & & \\
\texttt{GALFIT} & 30.5 & 105 \\
\texttt{GALFITM} & 35.4 & 178 \\
\bottomrule
\end{tabular*}
\end{table}

As one can see from the table, single-thread performance of \texttt{IMFIT-MW} is comparable to \texttt{GALFITM}. Moreover, \texttt{IMFIT-MW} can benefit from multithreading: model image creation is parallelized between filters which introduce little overhead (which can be seen as the increase of CPU time), but decreases the overall time by a factor of a few.

For my specific benchmarks, \texttt{IMFIT-MW} is even faster both in time and number of iterations. For NGC~3367, \texttt{IMFIT-MW} and \texttt{GALFITM} do 28 and 52 iterations, and for NGC~4391 these numbers are 32 and 63, respectively. In the same time, the resulting models are almost identical, and the difference in final $\chi^2$ is less than 0.1\%, in both cases being a little better in \texttt{IMFIT-MW}. Note that, per iteration \texttt{GALFITM} is equal or even faster than \texttt{IMFIT-MW}. It is possible that different parametrization of hyperparameters in \texttt{IMFIT-MW} makes them more independent from each other, which may contribute to simplifying the fit process and decreasing number of iterations. For example, parameters related to brightness have highest possible degree and any combination of single-band parameters is possible. However, in \texttt{GALFITM} they are parametrized with Chebyshev polynomial coefficients, where change of each one affects all single-band parameters, whereas \texttt{IMFIT-MW} parametrizes all single-band parameters directly, so each one influences the brightness distribution only in one filter. Different optimization algorithms may also contribute to this difference: benchmarking was done with \texttt{IMFIT-MW} using trust region reflective algorithm (\texttt{method = 'least\_squares'} in \texttt{lmfit}). This also resulted to increased RAM usage for internal images by \texttt{IMFIT-MW}.

Regarding single-band performance, \texttt{IMFIT-MW} falls behind multithreaded \texttt{IMFIT} without multiprocessing as only per-band parallelization is implemented, but their single-threaded performance is compatible. Notably, for NGC~3367 \texttt{IMFIT} requires more iterations than \texttt{lmfit} (32 vs 20), and for NGC~4391 the situation is opposite (14 and 24 iterations); this means, times per iteration are similar and overhead is not significant. In both cases, programs produce the same result and the relative difference of their $\chi^2$ is less than 10$^{-6}$.

Multiwavelength fitting takes more time than single-band fits for all filters combined, but this is unavoidable. The number of hyperparameters is inherently larger than the number of wavelength-dependent parameters, whereas the former are used to fit multiwavelength data and the latter are relevant to single-band fitting.

\subsection{Validation}
\label{sec:validation}

\begin{figure}
	\centering
	\includegraphics[width=.9\columnwidth]{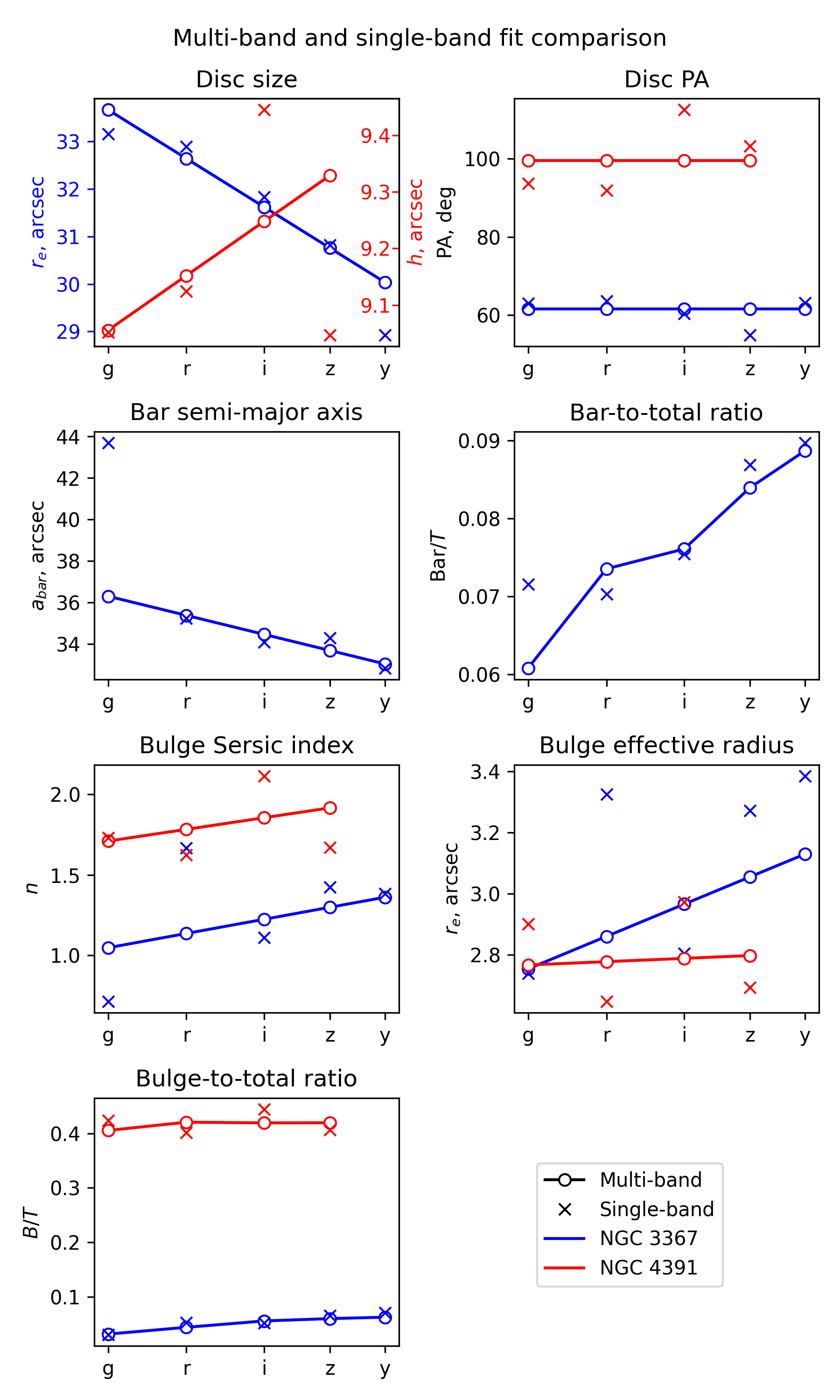}
	\caption{A comparison between multiwavelength and single-band model parameters behaviour with wavelength. Filters labels correspond to their wavelengths. Lines represent wavelength-dependent parameters obtained with multiwavelength fitting, whereas crosses are single-band results. Blue symbols represent the result for NGC~3367, and red ones are for NGC~4391.}
	\label{fig:validation}
\end{figure}

Now I present a validation by comparing multiwavelength and single-band fitting. Specifically, for both NGC~3367 and NGC~4391 and for each filter, I ran single-band fitting with \texttt{IMFIT} and compared the resulting models with multiwavelength models at each corresponding wavelength, produced by \texttt{IMFIT-MW}. In Fig.~\ref{fig:validation}, I demonstrate the behaviour of some important parameters and properties of models with wavelength.

For these two test galaxies, the parameters obtained from the multiwavelength fits are consistent with the trends measured independently using multiple single-band fits. Values related to component sizes or S{\'e}rsic indices follow the general trend that can be traced by single-band measurements, but outliers do not appear there. Bar- and bulge-to-total ratios do not deviate from single-band measurements as well. Position angle of the disc become constant and generally consistent with single-band values, removing physically implausible scatter between filters.

Note that NGC~3367 has complex structure in the centre, where one could easily expect some degree of degeneracy between bar, bulge and unresolved point source, and multiple local minima in the parameter space may present. Naturally, one can see jumps in bar and bulge sizes and bulge S{\'e}rsic index, especially in $gri$ filters. In contrast, with multiwavelength fitting linear parametrization of these quantities disallows abrupt changes between adjacent filters and therefore forces a model to behave consistently between filters \citep[compare with][]{Vika2013}. This may be particularly important when some filters have poor signal-to-noise ratio: without multiwavelength coupling, some of their fine parameters may be highly uncertain.

\subsection{IMFIT-specific model examples}

In this section, I demonstrate a few more examples of fitting multiwavelength models which employ custom and \texttt{IMFIT}-specific functions. The functions under discussion do not have direct \texttt{GALFIT} counterparts, so this usage example is unique to \texttt{IMFIT-MW}. All these functions are included to my modification of \texttt{PyImfit} (\url{https://github.com/IVChugunov/pyimfit}).

\begin{figure}
	\centering
	\includegraphics[width=.9\columnwidth]{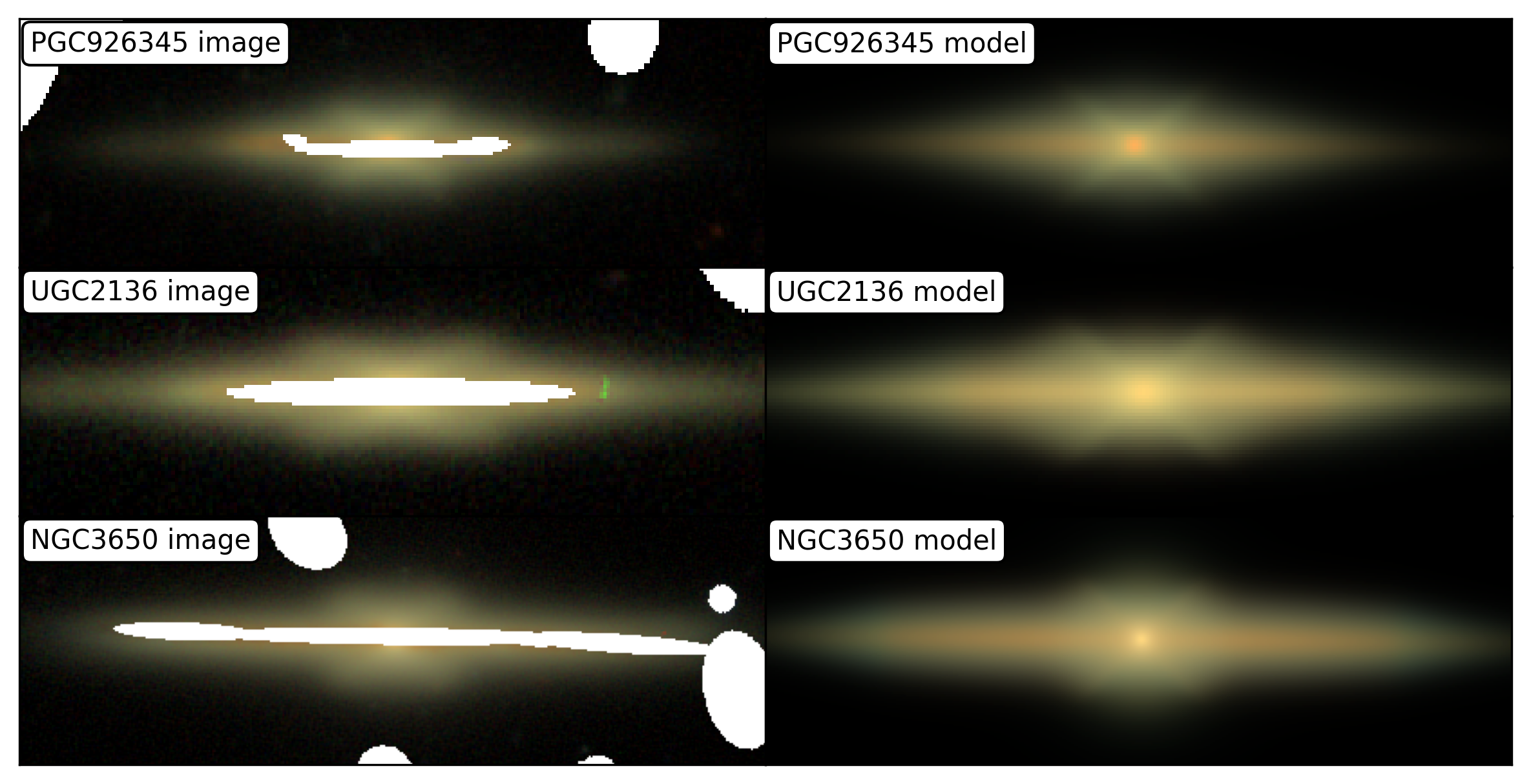}
	\caption{Images of three edge-on galaxies with X-structures and their multiwavelength models obtained with \texttt{IMFIT-MW}. RGB pictures are produced with $z$, $r$ and $g$ images using~\citet{Lupton2004} scaling.}
	\label{fig:BPS_models}
\end{figure}

The edge-on orientation of galaxies is useful for studying vertical structure of the discs, and projection effects associated with it improve detectability of disc truncations~\citep{Pohlen2004, Mosenkov2018}. As the first example, I select three edge-on galaxies from the sample of \citet{Savchenko2026} with the visually most prominent X-structures (which are simply bars observed in edge-on galaxies). For these objects I used DESI Legacy images: for two galaxies, $griz$ filters are available and for one --- only $grz$. I fit models which include not only bulge and disc representation, but also a custom B/PS bulge function from~\citet{Smirnov2026}. Two out of three galaxies also exhibit disc breaks, and in these cases discs are modelled using another custom function\footnote{\texttt{IMFIT} already features \texttt{BrokenExponentialDisk3D} which performs line-of-sight integration for disc with such break and arbitrary orientation, but it is computationally expensive. There is no analytical representation for edge-on orientation similar to \citet{vanderKruit1988} for disc without break where radial distribution is expressed through Bessel $K_1$ function. However, the vertical component of brightness distribution can be separated from radial one (which appears in LOS integration), and therefore one may calculate LOS integral only for a limited grid of radii, and interpolate for all pixels. Also, outside the break radius, where LOS does not cross break, the analytical solution of \citet{vanderKruit1988} remains applicable.}. In Fig.~\ref{fig:BPS_models}, I show the images and models I obtained for those galaxies. The subsample of edge-on galaxies with X-structures not complete and their modelling was done as a proof of concept, however more thorough analysis and interpretation may be done in future.

\begin{figure}
	\centering
	\includegraphics[width=.9\columnwidth]{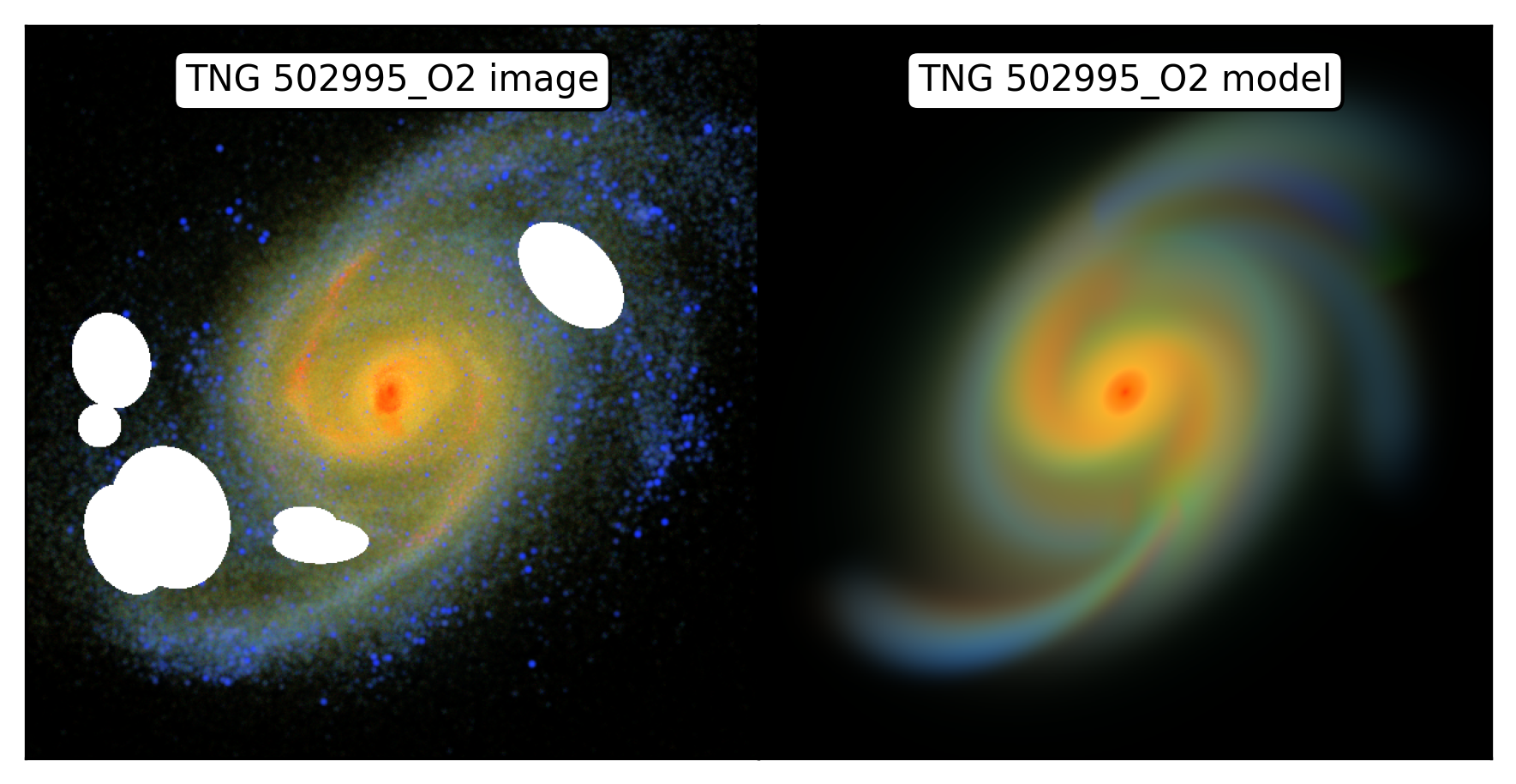}
	\caption{Image and multiwavelength model of simulated galaxy from TNG50 ID 502995 with orientation O2 (see \citealt{Baes2024}) including spiral structure, obtained with \texttt{IMFIT-MW}. RGB pictures are produced with $K_\text{S}$, $i$ and $u$ images using~\citet{Lupton2004} scaling; due to the selection of filters, colours look quite unnatural.}
	\label{fig:TNG_spiral_model}
\end{figure}

Another example involves modelling the spiral structure using~\citet{Chugunov2024, Chugunov2025b} model. TNG cosmological simulation~\citep{Nelson2019} data features multiple spiral galaxies, and the specific one that was modelled, has ID 502995 in TNG-50. This galaxy is present in TNG50-SKIRT Atlas \citep{Baes2024} where multiwavelength images are simulated for a number of galaxies, observed from different directions, using radiative transfer modelling. I constructed a multiwavelength model based on images in FUV, NUV, $u$, $i$ and $K_\text{S}$ filters which is shown in Fig.~\ref{fig:TNG_spiral_model}. A few models of simulated spiral galaxies, including ID 502995 which was demonstrated in this section, will be discussed in the following study (Zozulia et al., in prep.) and used to measure spatial offsets of spiral arms between different filters.

\section{Discussion and conclusions}

In this work, I have developed and presented \texttt{IMFIT-MW}: an extension to \texttt{IMFIT}, a program for photometric decomposition. The primary feature of \texttt{IMFIT-MW} is that it can work simultaneously with multiple images of a galaxy, allowing one to control wavelength dependence of photometric parameters. 

\texttt{IMFIT-MW} scope is similar to that of \texttt{GALFITM}, and both programs rely on the similar ideas and methods. Nevertheless, \texttt{IMFIT-MW} has some distinctive features that justify creation of the separate program, just like \texttt{IMFIT} and \texttt{GALFIT} are both in use nowadays.

First, \texttt{IMFIT-MW} is designed to be external program to \texttt{IMFIT} written in \texttt{Python} rather than a modification of its source code. As a result, \texttt{IMFIT-MW} is relatively lightweight and, because it is open source, easy to modify further by anyone for whatever features will be required in future. This also highlights the internal advantage of \texttt{IMFIT} modularity: one can access its functions separately and implement even those features for which original \texttt{IMFIT} was not designed.

For the benchmarks I considered above, \texttt{IMFIT-MW} single-thread performance is comparable to that of \texttt{GALFITM}: the overhead from \texttt{Python} code and communication between \texttt{PyImfit} and \texttt{lmfit} is small. Moreover, \texttt{IMFIT-MW} can benefit from multithreading, reducing the clock time required to process data significantly. However, there is a caveat that standard \texttt{PyImfit} appears to be less optimized than \texttt{IMFIT} in some routines. To avoid these problems, I recommend installing a version of \texttt{PyImfit} with my own minor modifications which allows the declared performance to be achieved. This version is available on GitHub (\url{https://github.com/IVChugunov/pyimfit}).

Finally, the feature which was actually the reason to create \texttt{IMFIT-MW} is the unique qualities of \texttt{IMFIT} itself. As mentioned before, \texttt{IMFIT} allows user to include their own brightness distribution functions. Although \texttt{GALFIT} offers a great deal of generalization of functions using truncations, Fourier and bending modes and similar features, these approaches are fundamentally different and one cannot fully replace another.

However, there are some limitations. As \texttt{IMFIT-MW} is an external project, its performance is anyway limited compared to what could be achieved if this was a genuine \texttt{C++} program. \texttt{IMFIT-MW} cannot fully reproduce \texttt{IMFIT} behaviour on single-band data, unlike \texttt{GALFITM} with respect to \texttt{GALFIT}.

This method will be particularly useful for the studies of spiral arms in galaxies, and in fact, a program that was developed into \texttt{IMFIT-MW} was initially created specifically for the related study (Zozulia et al., in prep.). The question of the nature of spiral structure remains unsolved, in part because different mechanisms of spiral structure formation are difficult to distinguish in observations \citep[see a review by][]{Sellwood2022}. One of the promising methods relies on the fact that spiral structures of different nature may have (or not have) different offsets between tracers of spiral structure such as gas, star formation and stellar population of different ages \citep[see][]{Tamburro2008, Kostiuk2024}. There were multiple attempts to measure such offsets, and some of them involved decomposition with spiral arms~\citep{Marchuk2024, Kostiuk2025}. However, the fundamental difficulty with it was that model spiral arms were often fitted to different features of spiral pattern in different filters, with models appearing dissimilar enough to conclude that offset measurement would not represent genuine physical offset. With \texttt{IMFIT-MW} and its ability to impose multiwavelength coupling between parameters, it became significantly easier to trace multiwavelength properties of the spiral pattern.

\section*{Acknowledgements}
This work is supported by funding of the Russian Science Foundation, grant № 25-22-00543.

I would like to thank Sergey S. Savchenko for the discussion of the idea underlying technical implementation of this paper. I would also like to acknowledge Peter Erwin, whose development of \texttt{IMFIT} and its design made this work possible. I am grateful to the anonymous referee for their comments which helped to improve the quality of the article.

\printcredits

\bibliographystyle{cas-model2-names}

\bibliography{cas-refs}

\end{document}